\documentclass[conference]{IEEEtran}
\IEEEoverridecommandlockouts
\usepackage{cite}

\usepackage{longtable}

\usepackage{array}
\usepackage{amsmath,amssymb,amsfonts}
\usepackage{algorithmic}
\usepackage{graphicx}
\usepackage{textcomp}
\usepackage{xcolor}
\def\BibTeX{{\rm B\kern-.05em{\sc i\kern-.025em b}\kern-.08em
    T\kern-.1667em\lower.7ex\hbox{E}\kern-.125emX}}
\begin{document}

\title{On the Integration of Course of Action Playbooks into Shareable Cyber Threat Intelligence\\
%Improving Integrated Adaptive Cyberspace Defense with Shareable Security Playbooks
%Integrating interoperable automated courses of action into cyber threat intelligence
%Automated Courses of Action in Cyber Threat Intelligence Sharing via Security Playbooks
%On the Role of Security Playbooks for Automated Courses of Action in Cyber Threat Intelligence Sharing
%Defensive and Adversarial Playbooks 
%improving the contextuality of cyber intelligence with shareable course of action playbooks
\thanks{This research has received funding from the European Health and Digital Executive Agency (HaDEA) under Grant Agreement No. INEA/CEF/ICT/A2020/2373266 for the JCOP project. In addition, part of this research has received funding from the European Union’s Horizon 2020 research and innovation program under Grant Agreement No. 830927 (CONCORDIA project) and 833418 (SAPPAN project). Furthermore, this research work was supported by the Research Council of Norway under Grant Agreement No. 303585 for the CyberHunt project.}
}

\author{\IEEEauthorblockN{Vasileios Mavroeidis}
\IEEEauthorblockA{\textit{University of Oslo}\\ Norway \\ vasileim@ifi.uio.no}
\and
\IEEEauthorblockN{Pavel Eis}
\IEEEauthorblockA{
\textit{Cesnet}\\
Czech Republic \\
eis@cesnet.cz}
\and
\IEEEauthorblockN{Martin Zadnik}
\IEEEauthorblockA{
\textit{Cesnet}\\
Czech Republic \\
zadnik@cesnet.cz}
\and
\IEEEauthorblockN{Marco Caselli}
\IEEEauthorblockA{
\textit{Siemens}\\
Germany \\
marco.caselli@siemens.com}
\and
\IEEEauthorblockN{Bret Jordan}
\IEEEauthorblockA{
\textit{Broadcom}\\
United States \\
bret.jordan@broadcom.com}
}

\maketitle

\begin{abstract}
Motivated by the introduction of CACAO, the first open standard that harmonizes the way we document courses of action in a machine-readable format for interoperability, and the benefits for cybersecurity operations derived from utilizing, and coupling and sharing course of action playbooks with cyber threat intelligence, we introduce a uniform metadata template that supports managing and integrating course of action playbooks into knowledge representation and knowledge management systems. We demonstrate the applicability of our approach through two use-case implementations. We utilize the playbook metadata template to introduce functionality and integrate course of action playbooks, such as CACAO, into the MISP threat intelligence platform and the OASIS Threat Actor Context ontology.
\end{abstract}


\section{Introduction}
To respond effectively and in a timely manner to cybersecurity events and incidents, defenders create and utilize security playbooks, also known as course of action playbooks. Security playbooks document processes and procedures that can guide, coordinate, and speed up security operations and incident response, ensure organizational policy and regulatory framework compliance, or purely automate security functions. 

CACAO is a security playbook standard developed by the OASIS Collaborative Automated Course of Action Operations Technical Committee (CACAO TC) \cite{cacao_tc}. CACAO defines a playbook schema and a taxonomy that standardize the way we create, document, and share course of action workflows \cite{cacao_spec_02}. A standard, machine-readable playbook schema like CACAO allows sharing playbooks across organizational boundaries and technological solutions seamlessly. Prior to CACAO, creating executable security playbooks relied on proprietary technologies that prevented their programmatic cross-utilization and sharing, also making it hard for users to compare generic playbooks and understand which offer the best models to leverage.

Cyber threat intelligence is evidence-based knowledge about adversaries and their operations. It increases defenders' threat situational awareness, supports decision making, and drives their proactive, active (i.e., active defense), and retroactive postures against threats. From a defender's point of view, a \textit{course of action} is a set of specific actions that could be taken to prevent or respond to an attack \cite{mavroeidis2017cyber}. From an attacker's point of view, a course of action is a set of specific actions pertinent to accomplishing the adversary's nefarious goals. Available machine-readable cyber threat intelligence representation and sharing standards and threat intelligence platforms provide textual course of action descriptions that are far away from being adequately structured and executable. With the introduction of CACAO, defenders can propel the contextuality and actionability of cyber threat intelligence by populating and sharing structured machine-readable course of action playbooks.

%This paper introduces a standard metadata template for integrating, collecting, managing, and sharing course of action playbooks as part of cyber threat intelligence.
In support of programmatically managing and sharing course of action playbooks, this paper introduces a standard metadata template to assist their integration into cyber-threat-intelligence-focused knowledge representation and knowledge management systems. We demonstrate the feasibility and the benefits derived from our standard templating approach by providing two use-case implementations. We show how security playbooks, such as CACAO, are integrated, managed, and shared using the MISP threat intelligence knowledge management, analysis, and sharing platform \cite{misp_paper_acm}\cite{misp_website} and a cyber threat intelligence ontology known as the Threat Actor Context ontology (TAC ontology) \cite{tac_tc}\cite{tac_onto_github}.

The rest of the paper is organized in the following way. Section \ref{sec:cacao-security-playbooks} introduces the CACAO security playbook standard. Section \ref{wrapper_schema} presents our proposed metadata template, which functions as a contextual integration layer between security playbooks and knowledge management systems. Sections \ref{sec:use-case-1} and \ref{sec:use-case-2} provide an overview of the MISP threat intelligence platform and the TAC ontology and discuss how our metadata template was utilized to introduce functionality for collecting, organizing, managing, and sharing security playbooks. Section \ref{sec:conclusion} concludes the paper.

\section{CACAO Security Playbooks}
\label{sec:cacao-security-playbooks}
\begin{figure}[htbp]
\includegraphics[width=0.70\columnwidth]{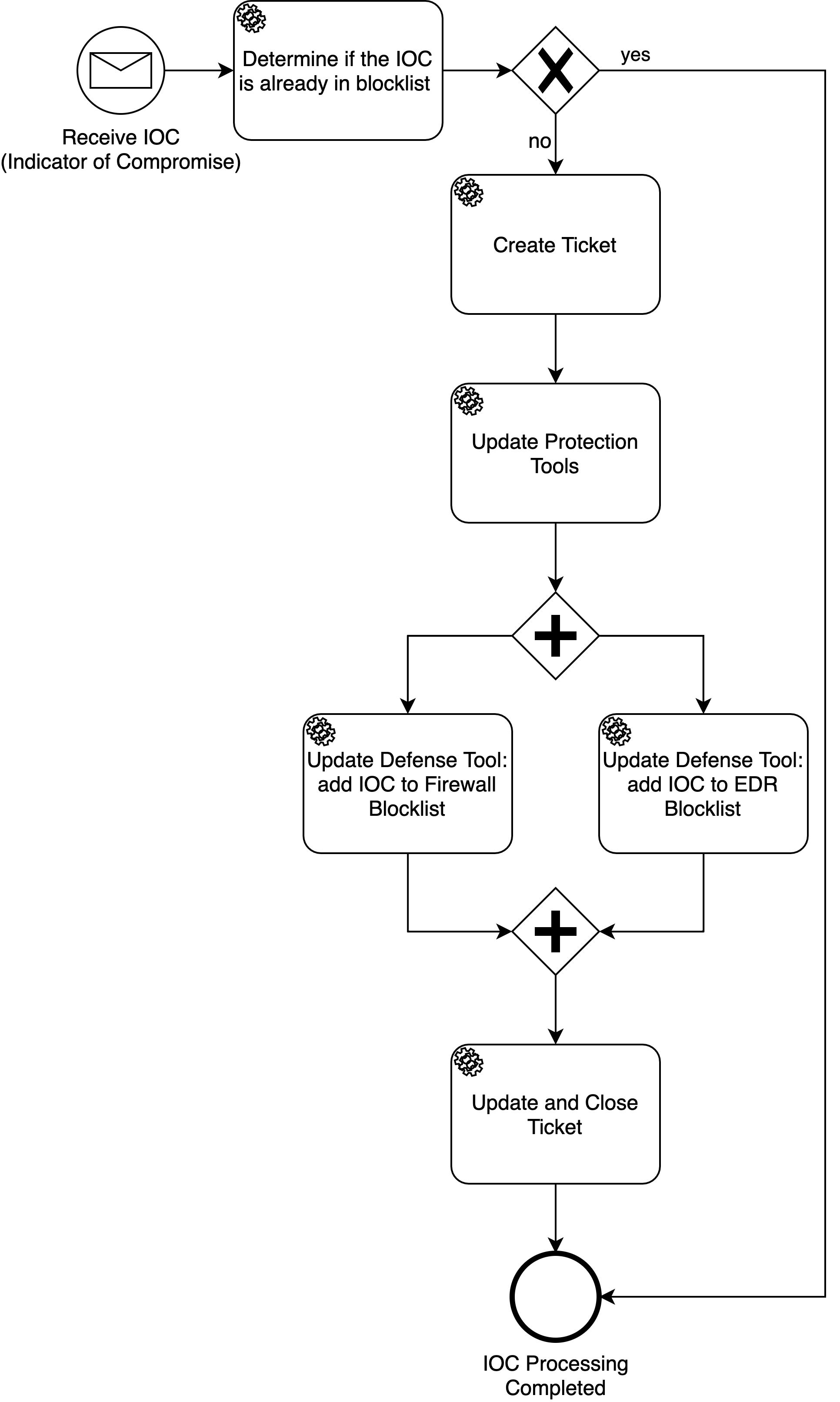}
\caption{Preventative security playbook (IoC handling).}
\label{abstract_playbook}
\end{figure}

Security playbooks are a fundamental component of Integrated Adaptive Cyberspace Defense (IACD) \cite{iacd2016}, guiding systems, subsystems, and human agents on how to interoperate to execute a course of action. Figure \ref{abstract_playbook} illustrates an abstract preventative playbook.  

A CACAO (Collaborative Automated Course of Action Operations) playbook is a workflow for security orchestration and automation represented in JavaScript Object Notation (JSON) that contains a set of steps to perform based on a logical process, similar to how Business Process Model and Notation (BPMN) defines a playbook for business processes. CACAO leveraged the design patterns of BPMN  to introduce a cybersecurity-specific JSON schema. 

The execution of a CACAO playbook may be triggered by an automated or manual event or observation or can be time-dependent (e.g., on a regular basis or periodically). 

\begin{figure}[htbp]
\begin{center}
\includegraphics[width=0.70\columnwidth]{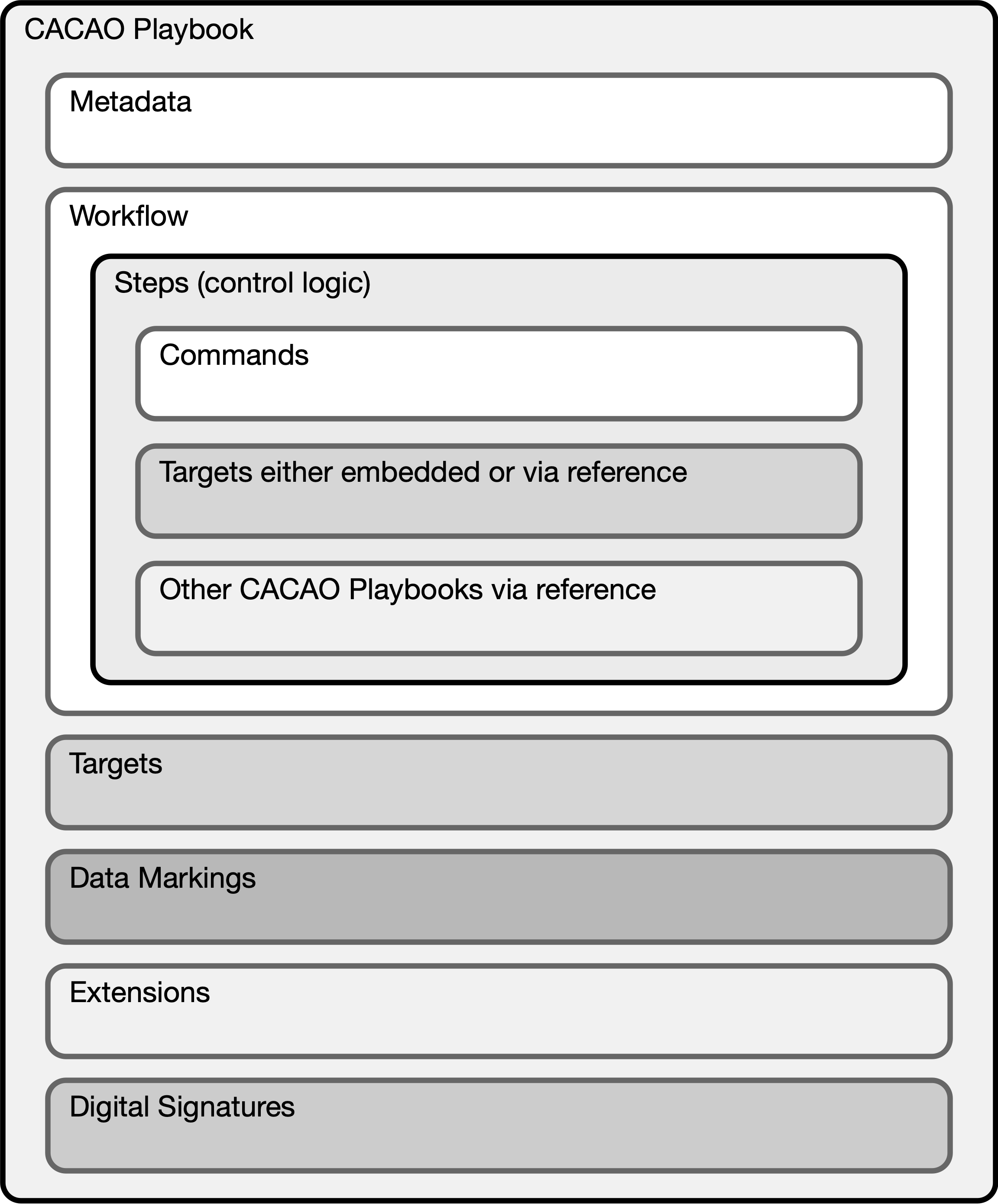}
\caption{Architecture and components of a CACAO security playbook.}
\label{cacao_architecture}
\end{center}
\end{figure}

At a high level, a CACAO playbook (Figure \ref{cacao_architecture}) comprises metadata and workflow steps that integrate logic to control the commands to be performed, a set of commands to perform, targets that receive, process and execute the commands, data markings that specify the playbook's handling and sharing requirements, and extensions that allow to granularly introduce additional functionality. Furthermore, for integrity and authenticity, CACAO playbooks can be digitally signed. The signature design supports both including the signature in the playbook itself and storing or releasing it separately as a detached signature.

CACAO playbooks can support multiple functions of cyberspace defense, such as threat hunting, detection, investigation, prevention, mitigation, remediation, or attack emulation. 

CACAO playbooks, among other command types, can encapsulate OpenC2 \cite{mavroeidis_openc2, openc2}, Sigma \cite{sigma}, and Kestrel \cite{kestrel} commands to support interoperability at the actuator level, making the playbooks require minimal modifications to map to an organization's own environment. 

%\section{A Common Metadata Template for Managing and Sharing Structured Course of Action Playbooks}
\section{A Common Metadata Template for Integrating Course of Action Playbooks into Shareable Cyber Threat Intelligence}
\label{wrapper_schema}
To integrate security playbooks into structured and shareable cyber threat intelligence, create managed security playbook knowledge bases, and manage security playbooks programmatically based on a set of meaningful attributes, we propose a metadata template (Table \ref{tab3}) that provides different technological solutions with a uniform approach. Producers and consumers of security playbooks can utilize the metadata template to contextually enrich their playbooks with additional machine-processable information as they make them available into their knowledge representation frameworks and knowledge management systems. An up-to-date version of the metadata template is available online\footnote{https://github.com/Vasileios-Mavroeidis/coa-playbook-metadata}.

The metadata available in CACAO playbooks have greatly influenced the development of the metadata template presented in this paper. Consequently, the metadata of a CACAO playbook can map directly to the metadata of the proposed template. 

The proposed metadata template is not restricted to encapsulating only machine-readable and automated courses of action. Similarly, it can provide other playbook representation types with metadata, making them programmatically available and manageable, such as a document describing a robust non-automated course of action or a playbook graph (model), like the one provided in Figure \ref{abstract_playbook}. 

%Based on the metadata template, security playbooks can stay intact in their native form or be encoded, with our set of metadata wrapping them so that they can be seamlessly extracted and machine-operationalized.

%date times in cacao different standard?  RFC 3339-formatted timestamp vs ISO 8601 format
\begin{table}[htbp]
\caption{Metadata Template for Security Playbook Management}
\begin{center}
\setlength{\extrarowheight}{2pt}
%\begin{tabular}{|c|p{0.65\linewidth}|}
\begin{tabular}{| >{\centering\arraybackslash}m{1in} | >{\raggedright\arraybackslash}m{2in} |}
\hline
\textbf{Element} & \textbf{Description}\\
\hline

id & A value that uniquely identifies the playbook.\\
\hline

created & The time at which the playbook was originally created.\\
\hline

modified & The time that this particular version of the playbook was last modified.\\
\hline

revoked & A boolean that identifies if the playbook creator deems that this playbook is no longer valid.\\
\hline

creator & The entity that created this playbook. It can be, for example, a natural person or an organization. It may be represented using an id that identifies the creator.\\
\hline

valid\_from & The time from which the playbook is considered valid and the steps that it contains can be executed.\\
\hline

valid\_until & The time at which this playbook should no longer be considered a valid playbook to be executed.\\
\hline

description & An explanation, details, and more context about what this playbook does and tries to accomplish.\\
\hline

label & An optional set of terms, labels, or tags associated with this playbook (e.g., aliases of adversary groups or malware family/variant/name that this playbook is related to).\\
\hline

impact & From 0 to 100, an integer representing the impact the playbook has on the organization. A value of 0 means specifically undefined. Values range from 1, the lowest impact, to a value of 100, the highest. For example, a purely investigative playbook that is non-invasive would have a low impact value of 1. In contrast, a playbook that performs changes such as adding rules into a firewall would have a higher impact value.\\
\hline

severity & From 0 to 100, an integer representing the seriousness of the conditions that this playbook addresses. A value of 0 means specifically undefined. Values range from 1, the lowest severity, to a value of 100, the highest.\\
\hline

priority & From 0 to 100, an integer representing the priority of this playbook relative to other defined playbooks. A value of 0 means specifically undefined. Values range from 1, the highest priority, to a value of 100, the lowest. \\
\hline

organization\_type & The type of organization that the playbook is intended for. This can be an industry sector.\\
\hline

playbook\_type & The security-related functions the playbook addresses. A playbook may account for multiple types (e.g., detection and investigation). ['Notification', 'Detection', 'Investigation', 'Prevention', 'Mitigation', 'Remediation', 'Attack'] \\
\hline

playbook\_standard & The standard the playbook conforms to (e.g., CACAO). \\
\hline

playbook\_abstraction & {The playbook's level of abstraction. ['Template', 'Executable']}\\
\hline

playbook & The whole playbook in its native format (e.g., CACAO JSON). Security playbook producers and consumers use this property to share and retrieve playbooks.\\
\hline

playbook\_base64 & The whole playbook encoded in base64. Security playbook producers and consumers of playbooks use this property to share and retrieve playbooks. \\
\hline
\end{tabular}
\label{tab3}
\end{center}
\end{table}

\section{Use Case: Introducing Security Playbooks in MISP Threat Intelligence Platform}
\label{sec:use-case-1}
Allowing one organization's detection to become another's prevention via intelligence sharing is a powerful paradigm that can advance the overall security of organizations \cite{johnson2016guide}. Emerging cybersecurity-focused knowledge management systems, otherwise known as Threat Intelligence Platforms (TIPs), equip defenders with the ability to collect, store, process, analyze, and share actionable information and intelligence in meaningful times. 

MISP (Malware Information Sharing Platform) \cite{misp_paper_acm} is an open-source threat intelligence platform that has undergone considerable development and increased functionality since 2011. Organizations use MISP to exchange cyber threat intelligence among trusted groups so that they can stay threat-informed and support their decision-making processes.

The fundamental building block of MISP is a structured core format for representing threat information and ensuring interoperability between MISP instances and other threat intelligence platforms that utilize it \cite{misp_core_format}. One component of the core format is the ability to create objects that serve as a contextual bond between a list of attributes. The main purpose of an object is to represent complex structures that could not be described by single attributes. Each object is created using an object template and carries the metadata of the template used for its creation within.

The existing MISP \textit{course of action} object (Table \ref{tab1}) is used to describe in a textual form a measure taken to prevent or respond to an attack.

\begin{table}[htbp]
\caption{MISP Course of Action Object Template}
\begin{center}
\setlength{\extrarowheight}{2pt}

\begin{tabular}{| >{\centering\arraybackslash}m{1in} | >{\raggedright\arraybackslash}m{2in} |}
\hline
\textbf{Object attribute} & \textbf{Description}\\
\hline
cost & The estimated cost of applying the course of action. ['High', 'Medium', 'Low', 'None', 'Unknown'] \\
\hline
description & A description of the course of action. \\
\hline
efficacy & The estimated efficacy of applying the course of action. ['High', 'Medium', 'Low', 'None', 'Unknown'] \\
\hline
impact & The estimated impact of applying the course of action. ['High', 'Medium', 'Low', 'None', 'Unknown'] \\
\hline
name & The name used to identify the course of action.\\
\hline
objective & The objective of the course of action. \\
\hline
stage & The stage of the threat management lifecycle that the course of action is applicable to. ['Remedy', 'Response', 'Further Analysis Required'] \\
\hline
type & The type of the course of action. ['Perimeter Blocking', 'Internal Blocking', 'Redirection', 'Redirection (Honey Pot)', 'Hardening', 'Patching', 'Eradication', 'Rebuilding', 'Training', 'Monitoring', 'Physical Access Restrictions', 'Logical Access Restrictions', 'Public Disclosure', 'Diplomatic Actions', 'Policy Actions', 'Other'] \\
\hline
\end{tabular}
\label{tab1}
\end{center}
\end{table}

\begin{figure}[htbp]
\begin{center}
\includegraphics[width=.68\columnwidth]{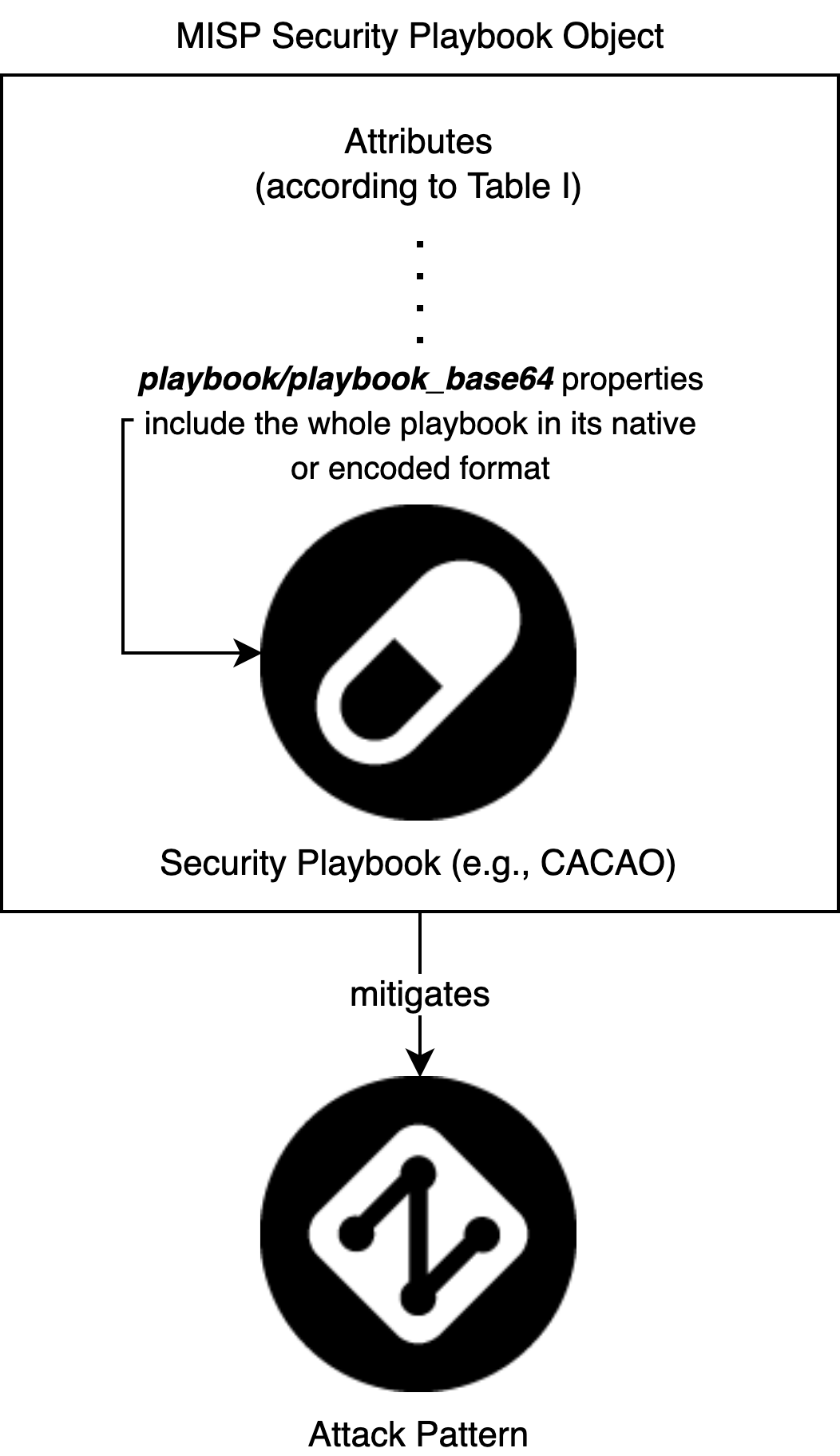}
\caption{Example illustration of MISP security playbook object.}
\label{fig:misp-cacao}
\end{center}
\end{figure}

Based on the common metadata template presented in Table \ref{tab3} (Section \ref{wrapper_schema}) in support of programmatically managing and sharing course of action playbooks, this research work introduced a conforming MISP security playbook object template that is the basis for creating shareable MISP security playbook objects. The security playbook object is a wrapper that encapsulates and makes available and shareable among MISP instances course of action orchestration workflows, such as CACAO security playbooks. A MISP security playbook object can make use of semantic relationships and link to other objects to improve the context around a playbook, such as the attack pattern a playbook mitigates or encodes to execute (attack emulation). An example is illustrated in Figure \ref{fig:misp-cacao}. The MISP security playbook object template is available at the official MISP GitHub repository\footnote{https://github.com/MISP/misp-objects/blob/main/objects/security-playbook/definition.json}.

%wrapper format for capturing both executable (i.e., automated course of action) and template security playbooks (i.e., course of action). In Section \label{}, we introduce a MISP security-playbook object based on the schema defined in Section \label{}. The new object allows MISP users and instances to share structured courses of action across organizational boundaries.

\section{Use Case: Introducing Security Playbooks in Threat Actor Context Ontology}
\label{sec:use-case-2}
Developed by the OASIS Threat Actor Context Technical Committee (TAC TC), the TAC ontology is an open-source modular knowledge representation framework that captures the rich context around adversaries into a structured machine-readable format. 

The core outline of the TAC ontology that defines the primary concepts, their relationships, and other semantics around the context of adversaries (in Web Ontology Language - OWL) is based on the STIX 2.1 standard, a model and a language to represent and exchange cyber threat intelligence (in JSON). The TAC ontology can operate as a knowledge management system that, based on its underlying representation schema, can be used to conduct intelligence analysis, fuse, store and share information, and perform logic-based information inference using a reasoner.

The framework's flexibility permits introducing and extending the existing core ontology with additional concepts and relationships, providing defenders with more extensive and integrated machine-readable cyber threat intelligence coverage and capability based on their organizations' needs. This also enhances the usability of STIX 2.1 by providing a bridge to other representations which can altogether harmonize with the TAC ontology. For example, in [2], the authors utilized a threat actor typology to characterize adversaries and, based on the adversaries defining characteristics, automatically infer their type (e.g., government cyberwarrior, cybercriminal). The ontology introduced in [2] demonstrates how their representation schema for threat actors could supplement or replace the threat actor concept/representation of the core TAC ontology as it is represented by the STIX 2.1 standard. 

The TAC ontology is fundamentally based on the STIX 2.1 standard and thus includes the concept of \textit{course of action}. According to the STIX 2.1 specification, the object supports basic use cases, like sharing prose courses of action, and does not support the ability to represent automated courses of action or contain properties to represent metadata about courses of action (Table \ref{tab2}). As annotated in the specification, future STIX 2 releases will introduce these capabilities.

\begin{table}[htbp]
\caption{STIX 2.1 Course of Action Object Template (Specific Properties)}
\begin{center}
\setlength{\extrarowheight}{2pt}

\begin{tabular}{| >{\centering\arraybackslash}m{1in} | >{\raggedright\arraybackslash}m{2in} |}

%\begin{tabular}{|c|p{0.65\linewidth}|}
\hline
%\textbf{Required Common Properties} & type, spec\_version, id, created, modified \\
%\hline
%\textbf{Optional Common Properties} & created\_by\_ref, revoked, labels, confidence, lang, external\_references, object\_marking\_ref, granular\_markings, extensions \\
%\hline
\textbf{Property Name} & \textbf{Description}\\
\hline

type & The value of this property MUST be course-of-action.\\
\hline

name & A name used to identify the Course of Action. \\
\hline

description & A description that provides more details and context about the Course
of Action, potentially including its purpose and its key characteristics. \\
\hline

action (reserved) & RESERVED – To capture structured/automated courses of action. \\
\hline

\end{tabular}
\label{tab2}
\end{center}
\end{table}

\begin{figure}[htbp]
\begin{center}
\includegraphics[width=1\columnwidth]{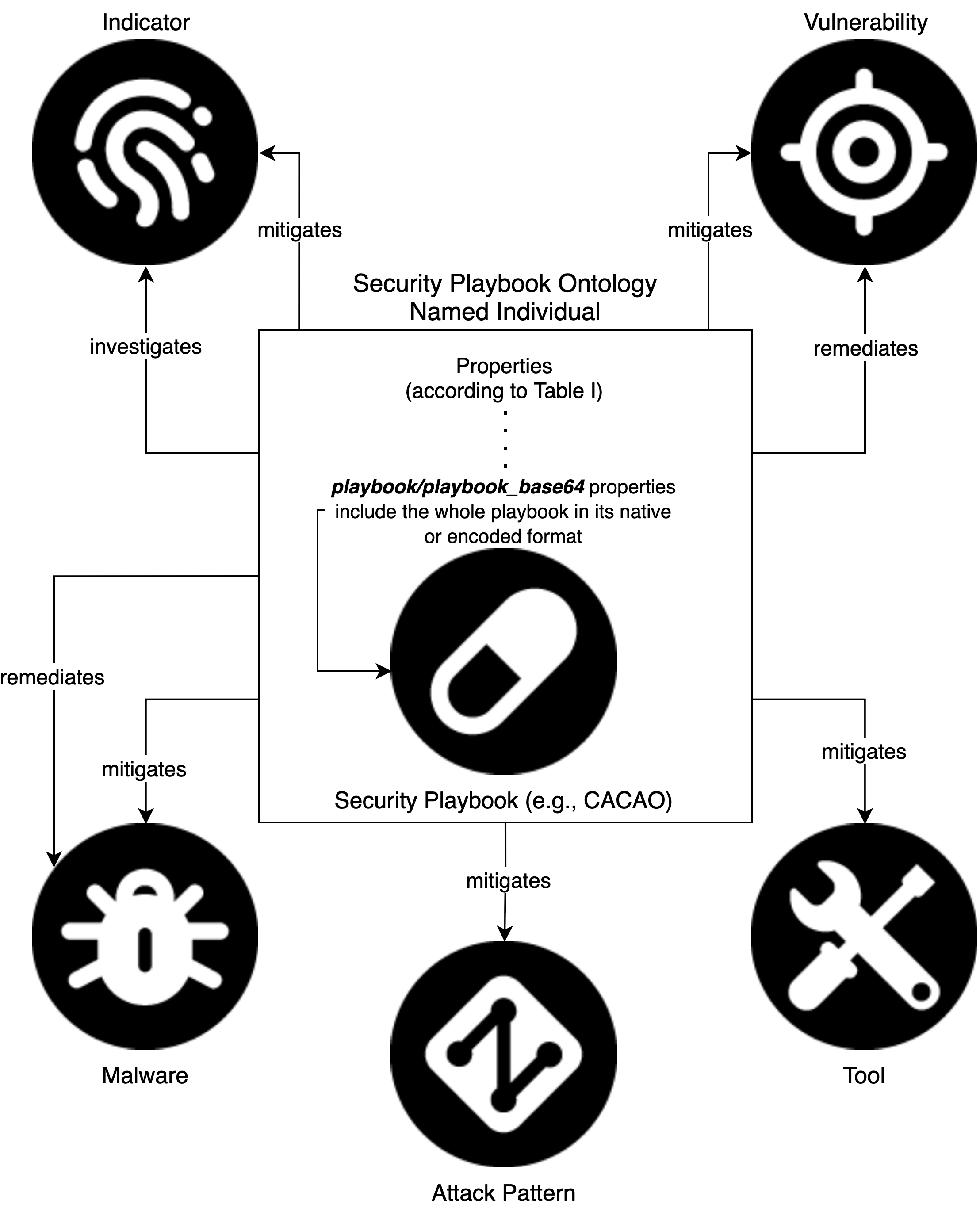}
\caption{Example illustration of TAC security playbook individual and its semantic relationships as specified on the STIX 2.1 standard.}
\label{stix_coa}
\end{center}
\end{figure}

Based on the common metadata template presented in Table \ref{tab3} (Section \ref{wrapper_schema}) in support of programmatically managing and sharing course of action playbooks, this research work introduced a conforming ontological representation in OWL. The proposed ontology can be integrated with other ontologies such as the TAC ontology to improve the context around adversaries with relevant security playbooks or can be utilized standalone to create a searchable (based on the properties defined in Table \ref{tab3}) knowledge base of security playbooks. The security playbook concept/ontology when integrated with the TAC ontology "inherits" the semantic relationships of the \textit{course of action} object as defined in STIX 2.1 (Figure \ref{stix_coa}). The ontology is maintained by the TAC TC\footnote{https://github.com/oasis-open/tac-ontology/blob/main/stix-semex/security-playbook}. 

\section{Conclusion}
\label{sec:conclusion}
While cyberspace defense continuously becomes more automated and adaptive, the need for sharing automated courses of action across organizational boundaries becomes apparent. CACAO is the first open non-proprietary standard that harmonizes the way we document executable courses of action. This paper presented a common metadata template that supports managing and integrating course of action playbooks, such as CACAO or other more abstract or proprietary, into knowledge management systems like threat intelligence platforms and threat-intelligence-focused knowledge representation approaches.\\

\section*{Acknowledgment}

The authors would like to thank Professor Audun Jøsang (University of Oslo), Jane Ginn (Cyber Threat Intelligence Network), Allan Thomson (Chief Architect Threat Defense Avast), and Francisco Luis de Andrés Pérez for providing feedback on the paper.

\IEEEpeerreviewmaketitle

%\clearpage
	\bibliographystyle{IEEEtran}
	\bibliography{bibliography}
	
\end{document}